\documentclass[english]{jpp}
\usepackage{lmodern}
\usepackage[T1]{fontenc}
\usepackage[latin9]{inputenc}
\usepackage{color}
\usepackage{babel}
\usepackage{amsmath}
\usepackage{amssymb}
\usepackage{graphicx}
\PassOptionsToPackage{normalem}{ulem}
\usepackage{ulem}
\usepackage[unicode=true,
 bookmarks=false,
 breaklinks=true,pdfborder={0 0 1},backref=false,colorlinks=true]
 {hyperref}
\hypersetup{
 hypertexnames=false,citecolor={blue},urlcolor={blue},linkcolor={blue}}

\makeatletter

\usepackage{float}
\usepackage{babel}
\usepackage{dcolumn}
\usepackage{bm}
\usepackage{color}

\makeatother

\begin{document}
\title{The dispersion and propagation of topological Langmuir-cyclotron waves in cold magnetized plasmas}
\author{Yichen Fu \aff{1,2} \corresp{\email{yichenf@princeton.edu}}, Hong Qin \aff{1,2} \corresp{\email{hongqin@princeton.edu}}}
\affiliation{\aff{1} Princeton Plasma Physics Laboratory, Princeton, NJ 08540 \aff{2} Department of Astrophysical Sciences, Princeton University, Princeton, NJ 08540}
\maketitle
\begin{abstract}
Topological Langmuir-Cyclotron Wave (TLCW) is a recently identified topological surface excitation in magnetized plasmas. We show that TLCW originates from the topological phase transition at the Langmuir wave-cyclotron wave resonance. By isofrequency surface analysis and 2D and 3D time-dependent simulations, we demonstrate that the TLCW can propagate robustly along complex phase transition interfaces in a unidirectional manner and without scattering. Because of these desirable features, the TLCW could be explored as an effective mechanism to drive current and flow in magnetized plasmas. The analysis also establishes a close connection between the newly instituted topological phase classification of plasmas and the classical CMA diagram of plasma waves. 
\end{abstract}

\section{Introduction}

Surface waves in plasmas have been extensively studied \citep{trivelpiece1959space,gradov1983linear,vladimirov1994recent} due to their wide range of applications \citep{moisan1979theory,ida1971analytical,lopez1979use,agranovich2012surface,kupersztych2004electron,ionson1978resonant}. Recently, a new class of surface modes in plasma produced by topological phase transition have been investigated \citep{parker2020topological,fu2021topological}. Here, we report studies on the dispersion and propagation of the newly identified topological surface mode called Topological Langmuir-Cyclotron Wave (TLCW) in cold magnetized plasmas.

Surface plasma waves are fluctuations propagating along the boundary between two regions. The energy of surface waves is usually concentrated at the boundary and attenuates rapidly in the normal direction. On the other hand, bulk (or body) waves are not localized in the boundary and propagate through plasmas. Bulk and surface waves were traditionally considered distinct waves and studied separately. It was realized in recent decades that there exists a connection between some surface waves and the topological properties of bulk waves. In the study of the integer quantum Hall effect \citep{thouless1982quantized}, it is found that the bulk states in the quantum Hall systems can be characterized by a topological invariant $n\in\mathbb{Z}$ called the Chern number. At the interface between a quantum Hall state and vacuum, chiral surface (edge) states exist \citep{halperin1982quantized} whose quantity is equal to the Chern number in the bulk state. Such a correspondence between bulk topological invariants and the number of chiral edge states can be generally summarized as the bulk-edge correspondence \citep{hasan2010colloquium}. The bulk topology and its relation with surface states have been widely studied not only in quantum systems \citep{hasan2010colloquium,bernevig2013topological,qi2011topological,armitage2018weyl}, but also in classical systems, such as those in photonics \citep{haldane2008possible,raghu2008analogs,ozawa2019topological,marciani2020chiral}, acoustics \citep{yang2015topological,wang2015topological,he2016acoustic}, mechanics \citep{kane2014topological,nash2015topological,huber2016topological} and fluid dynamics \citep{delplace2017topological,perrot2019topological,tauber2019bulk,souslov2019topological,venaille2021wave,zhu2021topology}.

Magnetized plasmas have been studied from the topological point of view as well. Cold plasmas in a constant magnetic field $B_{0}\hat{z}$ were found to have non-trivial topology. When variation in the $z$-direction can be ignored, i.e., $k_{z}=0$, Chern numbers for the X waves in 2D plasmas were calculated \citep{silveirinha2015chern,gangaraj2017berry} and related to surface waves via the bulk-edge correspondence \citep{silveirinha2016bulk}. When $k_{z}\neq0$, the non-trivial topology of the plasma as a 2D system were found \citep{parker2020topological,parker2021topological} and its topological phase diagram was established \citep{fu2021topological}. In particular, under-dense plasmas can be regarded as a Weyl semimetal \citep{gao2016photonic,yang2016one} and host a surface state with the Fermi-arc structure \citep{armitage2018weyl}. The topological matter properties of ideal MHD plasmas with magnetic shear \citep{parker2020nontrivial}, kinetic plasmas \citep{parker2021berry}, and surface waves within the continuous spectrum \citep{rajawat2022continuum} were also investigated.

The topological surface wave in the present study is called Topological Langmuir-Cyclotron Wave (TLCW), because it is localized at the interface between two plasmas in different topological phases separated by the phase transition at the resonance between the Langmuir wave and the cyclotron wave \citep{fu2021topological}. Based on a one-dimensional model, we study the dispersion of the TLCW using its isofrequency contours, from which the main physical properties of the TLCW, such as the unidirectional (chiral) propagation and the immunity of scattering, can be directly observed. The isofrequency contours also illustrate how the topology of the index-of-refraction surface of bulk plasma waves evolves when the TLCW exists. This establishes an interesting relationship between the topological classification by the Chern number and the well-known CMA diagram \citep{clemmow1955dependence,allis1959waves,allis1963waves} of plasma waves. Two- and three- dimensional time-dependent simulations have been performed for the linearized fluid equations. The TLCW are excited by a Gaussian source at a given frequency and propagates in various configurations, confirming the physical properties found in the analysis using the isofrequency contours. The momentum and angular momentum carried by the TLCW are also studied. Because the excitation and propagation of the TLCW are topologically protected, it could explored as a robust mechanism to inject energy and momentum into magnetized plasmas.

The paper is organized as follows. Section~\ref{sec:theory} briefly introduces the theoretical model and the topological matter properties of cold magnetized plasmas. The frequency range and isofrequency contours of the TLCW are given in Sec.\,\ref{sec:1D_eigen}. The connection between the isofrequency contours and the CMA diagram is addressed. Section~\ref{sec:numerical} describes the numerical algorithms and 2D and 3D simulation results of the TLCW.

\section{Theoretical model\label{sec:theory}}

This section describes the governing equations for dynamics in cold magnetized plasmas and briefly discusses the origin of the TLCW. We consider a cold stationary plasma with immobile ions, and the background magnetic field is constant, i.e., $\boldsymbol{B}_{0}=B_{0}\hat{z}$. Since there is no pressure in cold plasmas, any given plasma density profile $n(\boldsymbol{r})$ is an equilibrium. With proper renormalization, the linearized fluid equations can be written as \citep{stix1992waves,fu2021topological}: 
\begin{align}
 & \partial_{t}\boldsymbol{v}=-\omega_{\mathrm{p}}\boldsymbol{E}-\Omega\boldsymbol{v}\times\hat{z},\label{eq:basic1}\\
 & \partial_{t}\boldsymbol{E}=\nabla\times\boldsymbol{B}+\omega_{\mathrm{p}}\boldsymbol{v},\label{eq:basic2}\\
\  & \partial_{t}\boldsymbol{B}=-\nabla\times\boldsymbol{E},\label{eq:basic3}
\end{align}
where $\boldsymbol{v,}\boldsymbol{E},\boldsymbol{B}$ are perturbed velocity, electric and magnetic fields, $\omega_{\mathrm{p}}(\boldsymbol{r})=\sqrt{n(\boldsymbol{r})e^{2}/\epsilon_{0}m_{\mathrm{e}}}$ is the local plasma frequency, $\Omega=eB_{0}/m_{e}$ is the cyclotron frequency, $e>0$ is the elementary charge, $m_{\mathrm{e}}$ is the electron mass, and $\epsilon_{0}$ is the vacuum permittivity.

\begin{figure}
\centering 
\includegraphics[width=8cm]{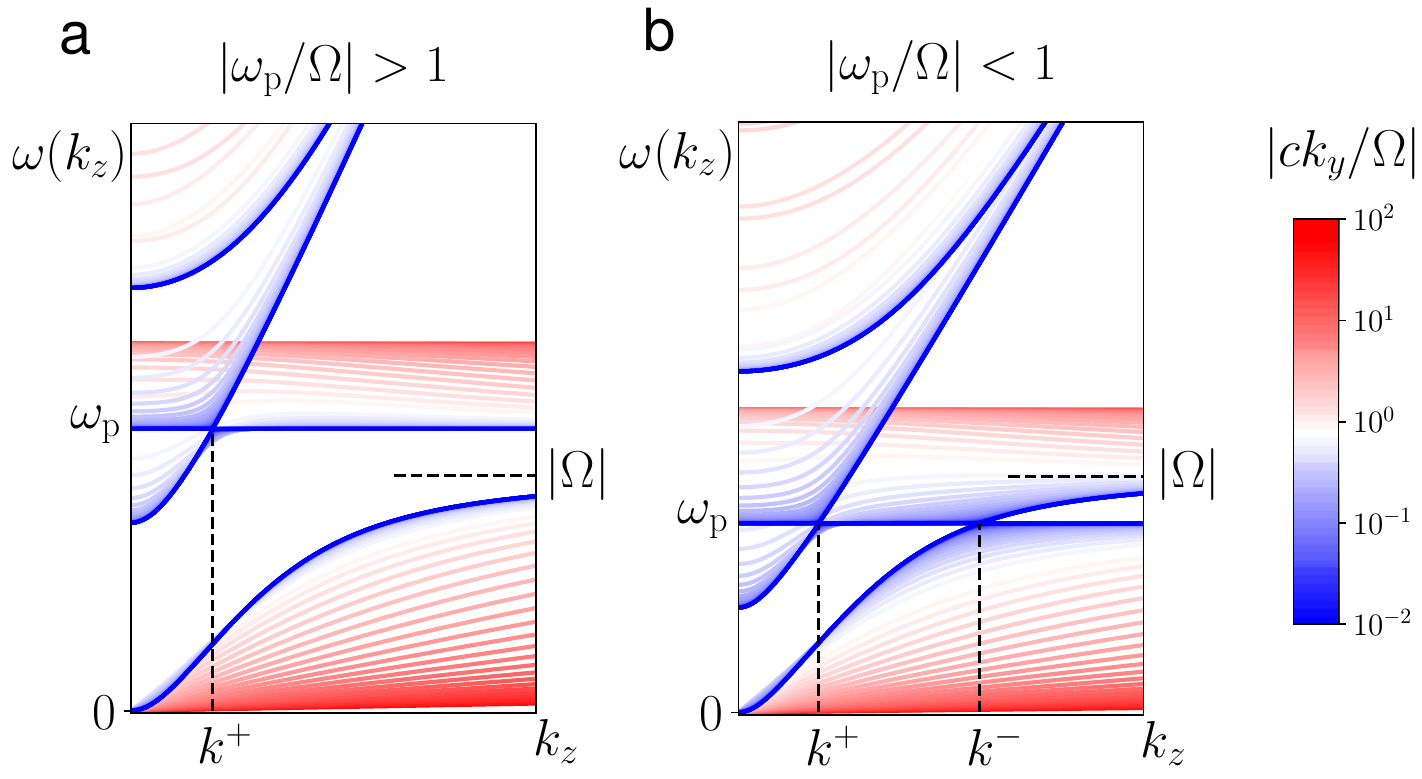} 
\caption{The dispersion relation $\omega_{n}(k_{z},k_{y})$ $(n=1,2,3,4)$ of cold plasma with immobile ions for (a) an overdense plasma, i.e., $|\omega_{\mathrm{p}}/\Omega>1|$, and (b) an underdense plasmas, i.e, $|\omega_{\mathrm{p}}/\Omega<1|$. Here, $k^{\pm}$ are the crossing points of $\omega_{1}$ and $\omega_{2}$, defined in Eq.\,(\ref{eq:kpm}). Different colors represent different values of $k_{y}$. Because the spectrum is also symmetric with respect to the rotation of $\boldsymbol{k}$ in the plane perpendicular to the magnetic field, $k_{x}$ is assumed to be zero. The dispersion relation $\omega(k_{y},k_{z})$ can be visualized by fixing $k_{z}$ and scanning all $k_{y}$ values.}
\label{fig:dispersion_relation} 
\end{figure}

In a homogeneous bulk plasma, for each eigenmode with frequency $\omega$ and wavenumber $\boldsymbol{k}$, Eqs.\,(\ref{eq:basic1})-(\ref{eq:basic3}) reduce to 
\[
H|\psi\rangle=\omega|\psi\rangle,
\]
where $|\psi\rangle=(\boldsymbol{v},\boldsymbol{E},\boldsymbol{B})^{\mathrm{T}}$ and 
\begin{align}
H(\omega_{\text{p}},\Omega,\boldsymbol{k}) & =\begin{pmatrix}-\mathrm{i}\Omega\hat{z}\times & -\mathrm{i}\omega_{\text{p}} & 0\\
\mathrm{i}\omega_{\text{p}} & 0 & -\boldsymbol{k}\times\\
0 & \boldsymbol{k}\times & 0
\end{pmatrix}\label{eq:sup:hamiltonian}
\end{align}
is a $9\times9$ Hermitian matrix. For each $\boldsymbol{k}$, the system has 9 eigenmodes 
\[
\omega_{n},|\psi_{n}\rangle,n=-4,-3,\cdots,3,4\thinspace.
\]
ordered by its value, i.e., $\omega_{i}\leq\omega_{j}$ for $i<j$. The spectrum is symmetric with respect to the real axis, i.e., $\omega_{-n}=-\omega_{n}$ and $\omega_{0}=0$. The dispersion relations of $\omega_{n}$ $(n=1,2,3,4)$ for over-dense and under-dense plasmas are plotted in Fig.\,\ref{fig:dispersion_relation}. Because the spectrum is also symmetric with respect to the rotation of $\boldsymbol{k}$ in the plane perpendicular to the magnetic field, $\omega_{n}$ is plotted only as functions of $k_{z}$ and $k_{y}$. The resonance between $\omega_{1}$ and $\omega_{2}$ happens at 
\begin{align}
k^{\pm}(\omega_{\mathrm{p}},\Omega):=\frac{\omega_{\mathrm{p}}/c}{\sqrt{1\pm\omega_{\mathrm{p}}/|\Omega|}}.\label{eq:kpm}
\end{align}

Topological surface modes might exist at the boundary between two bulk plasmas that share a common eigen-frequency gap and are in different topological phases. Consider two adjacent regions of plasmas with density $n_{1}$ and $n_{2}$, $n_{1}>n_{2}>0$, referred to as regions one and two, and assume that the interface between two regions lies in the $y$-$z$ plane. It has been demonstrated that \citep{fu2021topological} the TLCW exists if and only if the plasma parameters satisfy the inequality, 
\begin{align}
\frac{\omega_{\mathrm{p},1}}{|\Omega|}+\frac{\omega_{\mathrm{p},1}^{2}}{c^{2}k_{z}^{2}}>1>\frac{\omega_{\mathrm{p},2}}{|\Omega|}+\frac{\omega_{\mathrm{p},2}^{2}}{c^{2}k_{z}^{2}},\label{eq:condition}
\end{align}
where $\omega_{\mathrm{p},i}$ are the plasma frequencies in each region. Two regions of plasmas satisfying the inequality above are topologically different, or equivalently, there is a topological phase transition from one region to the other.

The condition was established by invoking the principle of bulk-edge correspondence in condensed matter physics for two bulk materials in different topological phases. The topological phases are characterized by the numerically calculated Chern numbers over the 2D space of perpendicular wavenumbers \citep{parker2020topological,fu2021topological}. Because the phase transition responsible for this particular surface mode is due to the resonance between the Langmuir wave and cyclotron wave, a proper name for it is Topological Langmuir-Cyclotron Waves (TLCW).

However, unlike the scenarios in condensed matters, where the wavenumber space is periodic, the wavenumber space for plasma waves and many other waves in classical media is topologically contractible. It is known that the topology of vector bundles over a topologically contractible base manifold is trivial. In addition, the Atiyah-Patodi-Singer index theorem \citep{Atiyah1976} of spectral flow, which is a rigorous mathematical statement for the bulk-edge correspondence, was only proved for the periodic wavenumber space. But for plasmas, the wavenumber space is not periodic in general. Recently, a rigorous analysis \citep{qin2022topological} has been developed using the tools of algebraic topology and a spectral flow index theorem formulated by Faure \citep{Faure2019} over $\mathbb{R}$-valued wavenumber space. The analysis confirms the existence of TLCW under condition (\ref{eq:condition}).

\section{The isofrequency contours of the TLCW\label{sec:1D_eigen}}

To quantitatively study the properties of the TLCW, we assume in this section that the background plasma density $n(\boldsymbol{r})$ is non-uniform only in the $x$-direction. The dispersion relation $\omega(k_{y},k_{z})$ of a homogeneous bulk can be displayed by two different methods. The first method is to plot $\omega(k_{y},k_{z})$ as a function of $k_{y}$ and $k_{z}$ as in Fig.\,\ref{fig:dispersion_relation}, which shows the range of band gaps and the location of the gap-closing. The other method to exam the level sets of the eigen-frequencies in the space of wavenumber, i.e., the isofrequency contours, which are more suitable for visualizing the group velocity. The familiar CMA diagram \citep{clemmow1955dependence,allis1959waves,allis1963waves} of bulk plasma waves can also be established as isofrequency contours, indicating an interesting connection between the CMA classification and the recently instituted topological phase classification of plasmas. In this section, we derive the frequency range where the TLCW exists and analyze the unique properties of the TLCW by drawing the isofrequency contours for both bulk and surface plasma waves in a one-dimensional model within the frequency range.

\subsection{The frequency range of the TLCW}

When the TLCW exists at the interface between two regions of different topological phases, it will fall into the common frequency gap of the bulk waves shared by the two regions. At the interface between two regions of different topological phases, a topological edge mode, also known as the spectral flow \citep{Faure2019,delplace2022berry,qin2022topological}, exists and it transits between the spectrum corresponding to the two regions. Specifically, in this paper, we define TLCW to be the topological edge mode in the frequency gap between the bulk spectrum. As shown later, mode within the bulk frequency gap is immune to back-scattering. Strictly speaking, frequencies of topological edge modes can overlap with the bulk spectrum, which is evident from the numerical results shown in Fig.\,\ref{fig:isofreq_nontrivial}. This is expected since the topological mode connects the gaped bands of the bulk spectrum. But to simplify the discussion with respect to the topological robustness, in the present context we choose to define TLCW in a narrow sense as the topological mode whose frequency is in the gap of the bulk bands. This restriction simplifies the discussion with respect to the topological robustness. We denote the frequencies of the first and second branches of bulk waves by $\omega_{1}(k_{y},k_{z})$ and $\omega_{2}(k_{y},k_{z})$, respectively. In each region, the frequency gap is between the top of the first branch $\max_{k_{y}}(\omega_{1})$ and the bottom of the second branch $\min_{k_{y}}(\omega_{2})$ \citep{fu2021topological}. Using this property, we obtain the possible frequency gap in each region, which can be written as 
\begin{align}
\min_{k_{z}}\left(\max_{k_{y}}(\omega_{1})\right)<\omega<\max_{k_{z}}\left(\min_{k_{y}}(\omega_{2})\right).\label{eq:possible_range}
\end{align}
Therefore, the common range given by the inequality above from the two regions determines the frequency range of possible TLCW. In other words, we need to solve inequality (\ref{eq:possible_range}) when TLCW exists, i.e., under the constraint of condition (\ref{eq:condition}). For simplicity, we define $k_{i}^{\pm}\equiv k^{\pm}(\omega_{\mathrm{p},i},\Omega)$ for $i=1,2$.

Firstly, consider region one with a higher density $n_{1}$. The wavenumber $k_{z}$ should satisfy the first inequality in condition (\ref{eq:condition}). If region one is overdense, i.e., $|\omega_{\mathrm{p,1}}/\Omega|>1$, the first inequality in condition (\ref{eq:condition}) holds for all $k_{z}$. From Fig.\,\ref{fig:dispersion_relation}(a), we can see that
\begin{align}
\min_{k_{z}\in\mathbb{R}}\left(\max_{k_{y}\in\mathbb{R}}(\omega_{1})\right)=\omega_{1}(0,0)=0,\quad\max_{k_{z}\in\mathbb{R}}\left(\min_{k_{y}\in\mathbb{R}}(\omega_{2})\right)=\omega_{2}(0,\infty)=\omega_{\mathrm{p},1}.\label{eq:extreme1}
\end{align}
If region one is underdense, i.e., $|\omega_{\mathrm{p,1}}/\Omega|<1$, the first inequality in condition (\ref{eq:condition}) gives $|k_{z}|<k_{1}^{-}$. Then from Fig.\,\ref{fig:dispersion_relation}(b) it is clear that 
\begin{align}
\min_{|k_{z}|<k_{1}^{-}}\left(\max_{k_{y}\in\mathbb{R}}(\omega_{1})\right)=\omega_{1}(0,0)=0,\quad\max_{|k_{z}|<k_{1}^{-}}\left(\min_{k_{y}\in\mathbb{R}}(\omega_{2})\right)=\omega_{2}(0,k_{1}^{-})=\omega_{\text{p},1}.\label{eq:extreme2}
\end{align}
Thus, for region one, the gap range is $0<\omega<\omega_{\text{p},1}$.

Next, for region two with a lower density $n_{2}$, the second inequality in condition (\ref{eq:condition}) shows that the plasma has to be underdense, i.e., $|\omega_{\mathrm{p,2}}/\Omega|<1$, and $|k_{z}|>k_{2}^{-}$. From Fig.\,\ref{fig:dispersion_relation}(b), we have
\begin{align}
\min_{|k_{z}|>k_{2}^{-}}\left(\max_{k_{y}\in\mathbb{R}}(\omega_{1})\right)=\omega_{1}(0,k_{2}^{-})=\omega_{\mathrm{p,2}},\quad\max_{|k_{z}|>k_{2}^{-}}\left(\min_{k_{y}\in\mathbb{R}}(\omega_{2})\right)=\omega_{2}(0,\infty)=|\Omega|.\label{eq:extreme3}
\end{align}
Thus, the gap range for region two is $\omega_{\mathrm{p,2}}<\omega<|\Omega|$. Combining these two ranges, we infer that the frequency range of possible TLCW is 
\begin{align}
\omega_{\mathrm{p,2}}<\omega<\min\big(|\Omega|,\omega_{\mathrm{p,1}}\big)\,.\label{eq:fre_range}
\end{align}

\subsection{Numerical demonstration of the frequency range}

\begin{figure}
\centering 
\includegraphics[width=8cm]{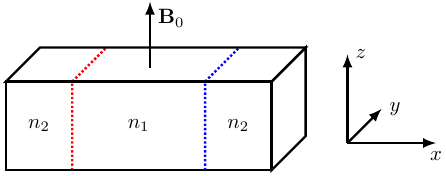} 
\caption{The equilibrium structure. The dotted red and blue lines are the boundaries between two regions.}
\label{fig:slab_geometry}
\end{figure}

After deriving the frequency range of TLCW in Eq.\,(\ref{eq:fre_range}), we now numerically demonstrate it using a one-dimensional model. We choose the density profile to be continuous, given by $n(x)=\frac{1}{2}(n_{1}-n_{2})\{\tanh[-(x-L)/\delta]+\tanh[(x+L)/\delta]\}+n_{2}$, where $n_{1}$ and $n_{2}$ are the plasma densities of regions one and two, respectively, and $L$ and $\delta$ are the location and width of the interface. The periodic boundary condition is used at $x=\pm2L$. The equilibrium configuration is sketched in Fig.\,\ref{fig:slab_geometry}. The width $\delta$ will only change the spectrum of bulk modes, but not affect the existence of TLCW, see \citep{qin2022topological} for a detailed discussion. In the present study, we used $L=40$ and $\delta=3$.

\begin{figure}
\centering 
\includegraphics[width=9cm]{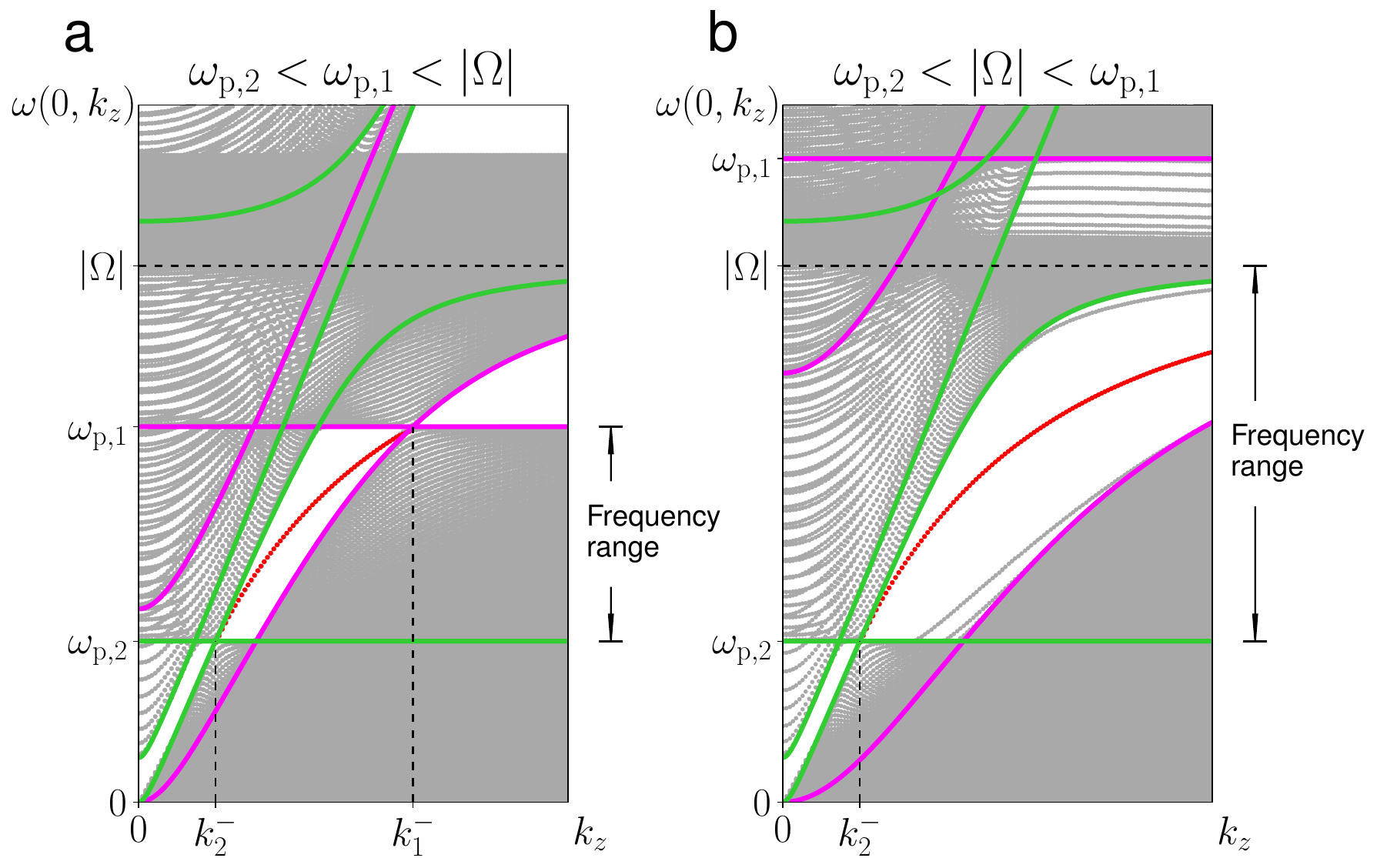} 
\caption{The dispersion relation $\omega(k_{y}=0,k_{z})$ in the non-uniform system. The green and magenta lines represent the bulk modes in region 1 and 2 at $k_{x}=k_{y}=0$; the red lines represent the TLCW; the gray curves represent the rest modes. The frequency ranges where TLCW exists are highlighted on the right of each plot. }
\label{fig:fermi_arc}
\end{figure}

Since all the extreme values of $\omega_{1,2}$ are reached at $k_{y}=0$ in Eqs.\,(\ref{eq:extreme1})-(\ref{eq:extreme3}), we show the numerically calculated dispersion relation of the non-uniform system $\omega(k_{z})$ at $k_{y}=0$. Two different cases are shown in Fig.~\ref{fig:fermi_arc}, which are adapted from \citep{fu2021topological}. The TLCWs are shown by the red curves, while the gray curves show all other modes. The green and magenta curves represent the bulk modes when $k_{x}=k_{y}=0$. In Fig.~\ref{fig:fermi_arc}a, both region 1 and 2 are underdense. We see that TLCW only exists when $k_{2}^{-}<k_{z}<k_{1}^{-}$, and its frequency range is $\omega_{\mathrm{p,2}}<\omega<\omega_{\mathrm{p,1}}$. In Fig.~\ref{fig:fermi_arc}b, region 1 is overdense while region 2 is underdense. Now TLCW exists when $k_{z}>k_{2}^{-}$ within the frequency range $\omega_{\mathrm{p,2}}<\omega<|\Omega|$. Notice that the frequency of TLCW converges to $|\Omega|$ when $k_{z}\to\infty$. Fig.~\ref{fig:fermi_arc} confirms the frequency range derived in Eq.\,(\ref{eq:fre_range}).

\subsection{The isofrequency contours from 1D model\label{sec:iso_contours}}

After determining the frequency range of possible TLCW, we numerically calculate the isofrequency contours for both the bulk and the surface waves. To draw the isofrequency contours of the system, the dispersion relation $\omega(k_{y},k_{z})$ and the eigenmodes are calculated using the method introduced in Ref.\,\citep{fu2021topological}.

\begin{figure}
\centering 
\includegraphics[width=8cm]{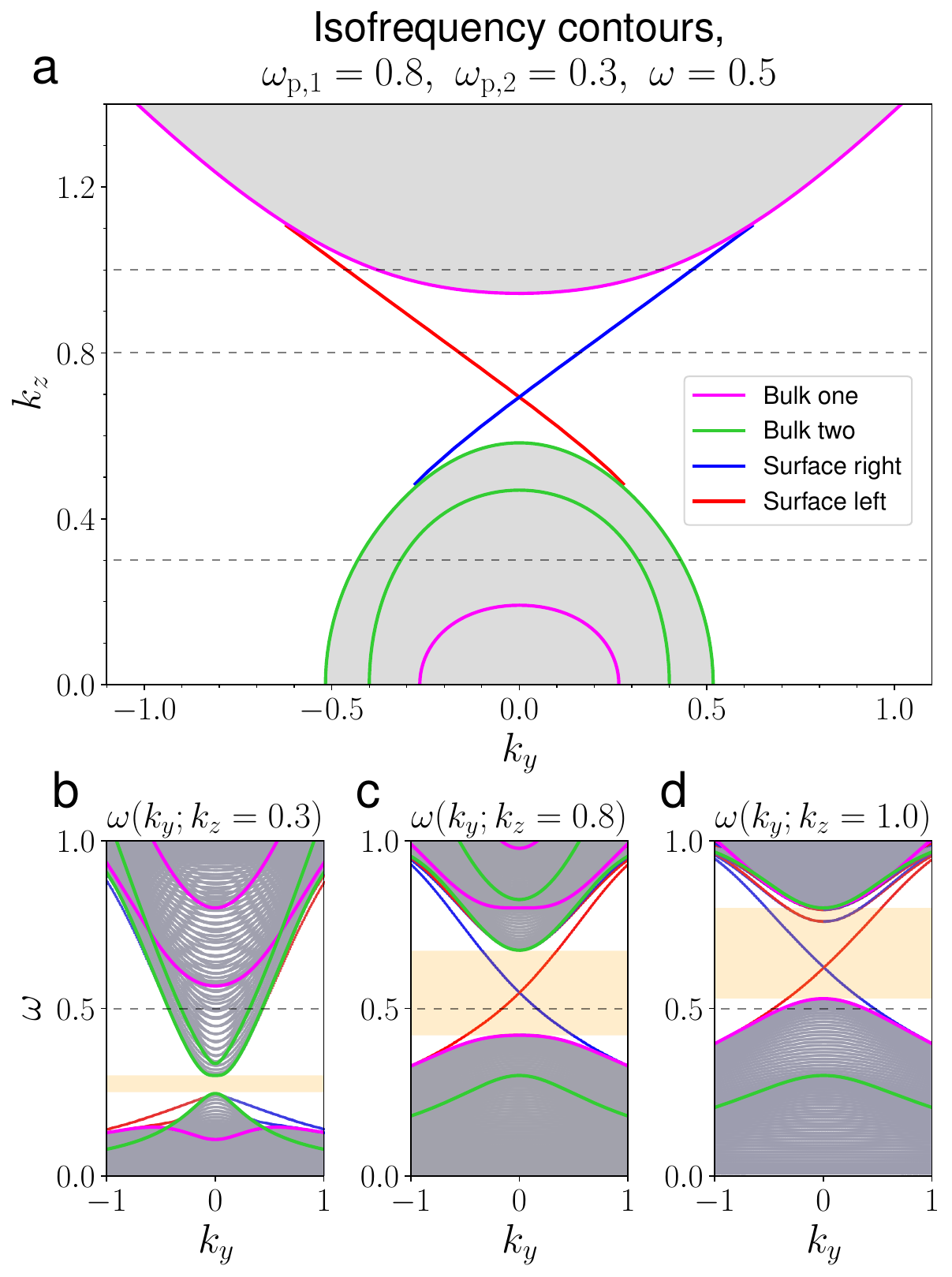} 
\caption{The dispersion relations for the system with $\omega_{\mathrm{p,1}}=0.8$ and $\omega_{\mathrm{p,2}}=0.3$. The magenta and green lines represent the isofrequency contours of bulk waves in regions one and two, assuming $k_{x}=0$. The blue and red lines represent the surface waves on the left and right boundaries. The gray areas and lines represent the bulk waves with $k_{x}\protect\neq0$. (a) The isofrequency contours at frequency $\omega=0.5$. The dashed lines represent the specific values of $k_{z}$. (b)-(d) The $\omega(k_{y})$ at $k_{z}=0.3,0.8$ and $1.0$, respectively. The dashed lines represent the location of $\omega=0.5$. The orange area are the frequency gap of bulk waves.}
\label{fig:isofreq_nontrivial} 
\end{figure}

Figure \ref{fig:isofreq_nontrivial} displays the isofrequency contours of the system when the TLCW exists. We choose cyclotron frequency to be $|\Omega|=1$, and both regions are underdense with $\omega_{\mathrm{p,1}}=0.8$ and $\omega_{\mathrm{p,2}}=0.3$. The frequency range of TLCW according to Eq.\,(\ref{eq:fre_range}) is $0.3<\omega<0.8$. The isofrequency contours at frequency $\omega=0.5$ are shown in Fig.\,\ref{fig:isofreq_nontrivial}a for $k_{z}>0$, and the $k_{z}<0$ part can be obtained from the symmetry condition $\omega(k_{y},k_{z})=\omega(k_{y},-k_{z})$. The dispersion relations $\omega(k_{y};k_{z})$ at $k_{z}=0.2,0.8$ and $1.0$ are shown in Fig.\,\ref{fig:isofreq_nontrivial}b-d for $\omega>0$, and the $\omega<0$ part can be obtained from symmetry condition $\omega(-k_{y},k_{z})=-\omega(k_{y},k_{z})$. In these plots, colored lines represent the surface modes or the bulk modes with $k_{x}=0$, while the gray lines/regions represent other bulk modes with $k_{x}=n\pi/2L$, where $n=\pm1,\pm2,\cdots$. 

For the isofrequency contours in Fig.\,\ref{fig:isofreq_nontrivial}a, the contours (green) for bulk waves in region two are both ellipsoids, while one of the contours (magenta) for region one is a hyperboloid. 
Well-defined surface waves at the left (red) and right (blue) boundaries can be found approximately around $0.5\leq k_{z}\leq1.1$, connecting the ellipsoidal and hyperboloidal contours. Outside this range of $k_{z}$, since the density profile is continuous, the eigenmodes of the surface waves decay slowly away from the boundaries into the bulk modes. A closer look can be taken at specific values of $k_{z}$. At $k_{z}=0.3$ from Figs.\,\ref{fig:isofreq_nontrivial}a and \ref{fig:isofreq_nontrivial}b, only bulk modes from region two exist. Near $k_{z}=0.8$, there is a gap for bulk modes in both Fig.\,\ref{fig:isofreq_nontrivial}a (around $0.58\leq k_{z}\leq0.95$) and Fig.\,\ref{fig:isofreq_nontrivial}c (around $0.42\leq\omega\leq0.67$). The surface modes that exist and fill-up such frequency gap are what we called TLCWs, predicted by the topological phase transition and the bulk-edge correspondence. At $k_{z}=1.0$, there exist both surface waves and bulk waves in region one. These surface waves are not topological surface waves because they are not in the frequency gap, though they are continuations of the topological surface waves into the frequency ragne of the bulk waves.

Some physical properties of the TLCW are directly observable from the isofrequency contours. Firstly, the TLCW is unidirectional. Since the group velocity $\boldsymbol{v}_{\mathrm{g}}=\partial\omega/\partial\boldsymbol{k}$ is perpendicular to the isofrequency contours, we can see that $v_{\mathrm{g},z}>0$ ($v_{\mathrm{g},z}<0$) for $k_{z}>0$ ($k_{z}<0$), and $v_{\mathrm{g},y}>0$ ($v_{\mathrm{g},y}<0$) on the left (right) boundary. Secondly, since by definition the TLCW refers to the topological modes in a frequency gap of bulk modes, it is immune to back-scattering when the surface is perturbed, at least when the density perturbation is a function of the $x$ and $y$ coordinates only, and the scale-length of the perturbation is much larger than the wavelength of the waves. Intuitively, this is because no other bulk wave exists for the surface waves to couple with in such a frequency gap. Although two surface waves exist in the frequency gap, they are spatially separated, i.e., one on the left and the other on the right boundary, and cannot interact. In section~\ref{sec:numerical}, these properties will be further verified by numerical simulations in both 2D and 3D.

\begin{figure}
\centering 
\includegraphics[width=8cm]{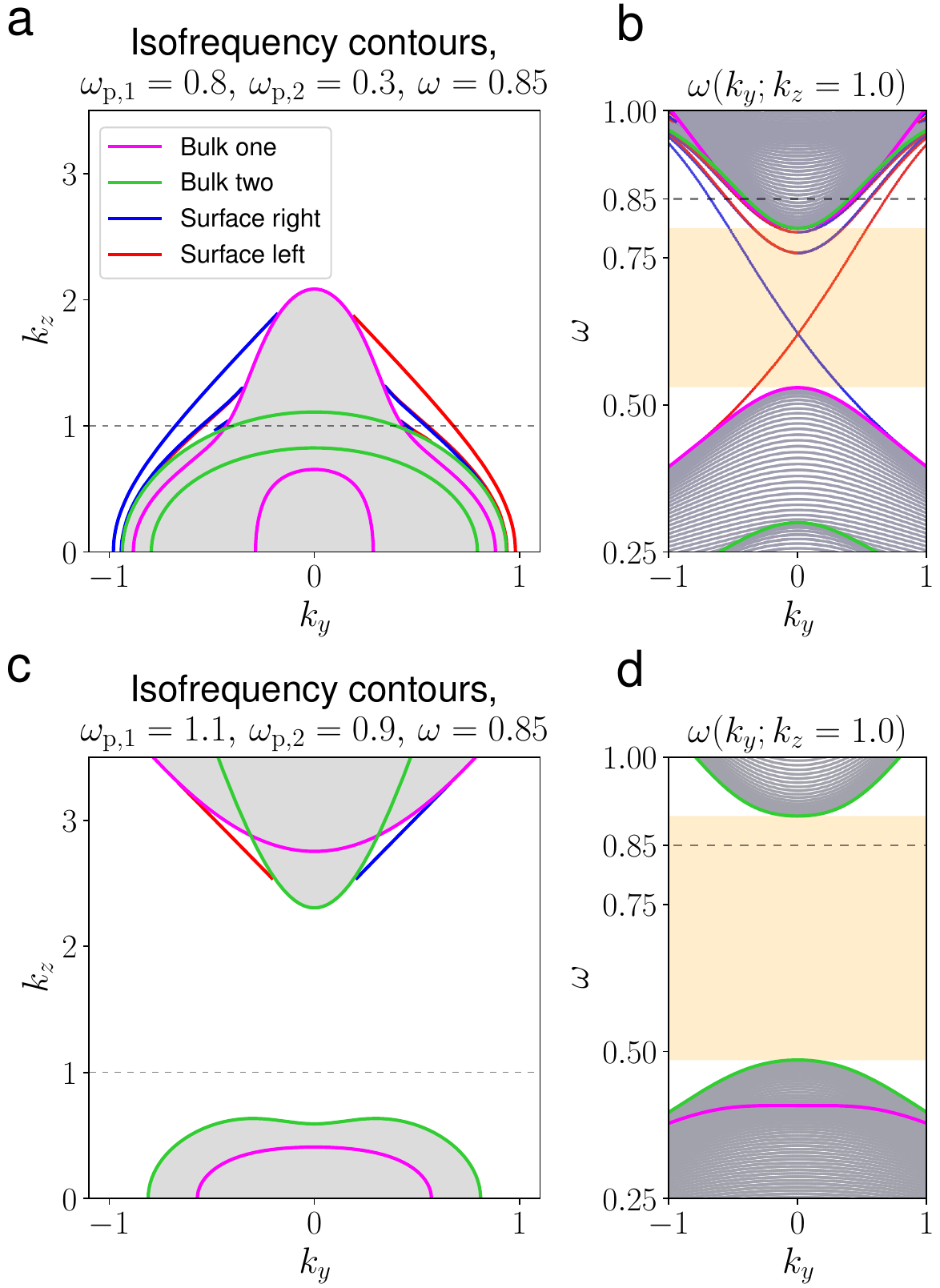} 
\caption{The isofrequency contours and dispersion relation $\omega(k_{y};k_{z})$ for (a) and (b) $\omega_{\mathrm{p,1}}=0.8$ and $\omega_{\mathrm{p,2}}=0.3$ , and for (c) and (d) $\omega_{\mathrm{p,1}}=1.1$ and $\omega_{\mathrm{p,2}}=0.9$. }
\label{fig:isofreq_trivial} 
\end{figure}

For comparison, two isofrequency contours at two frequencies where no TLCW exists are shown in Fig.\,\ref{fig:isofreq_trivial}. Again, we choose $\omega_{\mathrm{p,1}}=0.8$ and $\omega_{\mathrm{p,2}}=0.3$, and the frequency range for TLCW is $0.3<\omega<0.8$, if exists. The isofrequency contours at $\omega=0.85$ is plotted in Fig.\,\ref{fig:isofreq_trivial}a. Here, the isofrequency contours of the bulk waves for both regions are ellipsoid. There are multiple surface waves when $k_{z}<2.1$, some of which are the continuation of the TLCW, while others are just surface waves without topological origin. The dispersion relation $\omega(k_{y})$ at $k_{z}=1$ is shown in Fig.\,\ref{fig:isofreq_trivial}b, which shows that $\omega=0.85$ is not in the frequency gap of bulk modes. For the case in Fig.\,\ref{fig:isofreq_trivial}c, $\omega_{\mathrm{p,1}}=1.1$ and $\omega_{\mathrm{p,2}}=0.9$, and the frequency range for possible TLCW is $0.9<\omega<1$. At frequency $\omega=0.75$, the isofrequency contours in Fig.\,\ref{fig:isofreq_trivial}c show that both regions have an ellipsoidal and a hyperboloidal contours for bulk waves, and there is a bulk gap around $0.7<k_{z}<2.3$. As expected, there is no surface wave in the gap. At $k_{z}=1.0$, both Figs.\,\ref{fig:isofreq_trivial}c and \ref{fig:isofreq_trivial}d show that no surface mode exists at this $k_{z}$.

\subsection{Relation with the CMA diagram}

In the previous numerical calculation, we see that when the TLCW exists (Fig.\,\ref{fig:isofreq_nontrivial}a), the shapes of the isofrequency contours of bulk waves in regions one and two are different. On the contrary, when no TLCW exists (Figs.\,\ref{fig:isofreq_trivial}a and \ref{fig:isofreq_trivial}c), the shapes of contours of bulk waves in two regions are the same. Here, we demonstrate that this phenomenon originates from the relationship between the topological classification by Chern invariants and the well-known CMA classification of the plasma waves \citep{clemmow1955dependence,allis1959waves,allis1963waves}.

\begin{figure}
\centering 
\includegraphics[width=8cm]{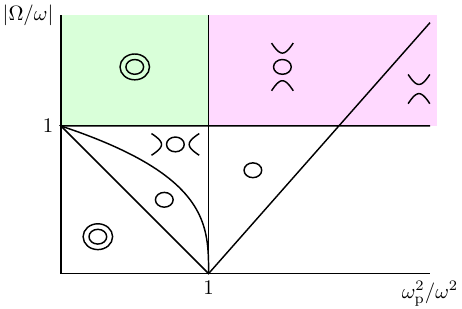} 
\caption{CMA diagram for cold plasmas with immobile ions \citep{clemmow1955dependence,allis1959waves,allis1963waves}. The shapes of the index-of-refraction surface are sketched in each region. When TLCW exists, region one must belong to the top right part filled by magenta, while region two must belong to the top left part filled by green.}
\label{fig:CMA} 
\end{figure}

In the study of homogeneous cold plasmas, waves at given frequencies can be classified based on the shapes of the wave normal surfaces ($\boldsymbol{v}_{\mathrm{p}}=\omega/c\boldsymbol{k}$), or equivalently the shape of the index-of-refraction surface ($\boldsymbol{n}=c\boldsymbol{k}/\omega$). We use the second method here since at a given frequency $\omega$ the shape of the index-of-refraction surface is the same as the isofrequency contours. Such surfaces have three possible shapes: ellipsoid, hyperboloid of two sheets, and hyperboloid of one sheet. The CMA diagram plots the different shapes of the index-of-refraction surface in the parameter space of the plasma density and magnetic field strength. A CMA diagram for cold plasmas with immobile ions is shown in Fig.\,\ref{fig:CMA}. In different regions of the CMA diagram, the shapes of the index-of-refraction surface may change. This is similar to the variation of Fermi surfaces, known as the Lifshitz phase transitions \citep{volovik2017topological} in condensed matter physics. This transition closely connects with the topological phase transition discussed in Sec.\,\ref{sec:theory}. The frequency range of the TLCW in Eq.\,(\ref{eq:fre_range}) can be equivalently written as 
\begin{align}
|{\Omega}/{\omega}|>1,\quad{\omega_{\mathrm{p,1}}^{2}}/{\omega^{2}}>1,\quad{\omega_{\mathrm{p,2}}^{2}}/{\omega^{2}}<1.\label{eq:fre_range2}
\end{align}
It turns out that on the CMA diagram, the inequalities in (\ref{eq:fre_range2}) also determine the possible shapes of bulk waves. Region one belongs to the two parts at the top right filled with magenta color, where one bulk wave is a hyperboloid. Region two belongs to the top left filled with green color, where all bulk waves are ellipsoid. Therefore, we find that the topological phase transition that generates the TLCW occurs simultaneously with the transition of the shapes of the index-of-refraction surfaces. In other words, the existence of TLCM implies different shapes of the bulk waves on the two sides of the interface and vice versa.

The CMA diagram includes all transitions of wave shapes, one of which corresponds to the TLCW. However, it is not known whether every transition of wave shape will lead to a topological edge mode. For magnetized cold plasmas, TLCW is the only known topological mode identified so far. Nevertheless, we suspect that similar mechanism might exist for other possible topological modes and the CMA diagram could be used as a guide for searching for these modes.

\section{Numerical simulations in 2D and 3D\label{sec:numerical}}

This section presents numerical simulation results of time-dependent propagation of the TLCW in two- and three-dimensions, based on the linearized incompressible fluid system specified by Eqs.\,(\ref{eq:basic1})-(\ref{eq:basic3}). We first introduce the numerical method adopted for time-dependent simulation and then discuss the simulation results and verify the key physics properties of the TLCW, including the unidirectional propagation and the immunity to scattering. The momentum and angular momentum of the waves will also be addressed. The time-dependent wave propagation in this section can be found in the supplementary movies.

\subsection{Numerical algorithms}

To simulate the dynamics governed by Eqs.\,(\ref{eq:basic1})-(\ref{eq:basic3}), we solve the momentum equation (\ref{eq:basic1}) using the Caylay transformation \citep{qin2013boris}, and solve the Maxwell equations (\ref{eq:basic2}) and (\ref{eq:basic3}) on the Yee grid \citep{yee1966numerical}. The 3D space is discretized by a $N_{x}\times N_{y}\times N_{z}$ grid, where the grid size in three directions are $\Delta x,\Delta y$ and $\Delta z$. In each direction, we define integer grid points and half-integer grid points, e.g., $x_{i}=i\Delta x$ and $x_{i+\frac{1}{2}}=(i+\frac{1}{2})\Delta x$. The time grid has a similar structure, i.e., $t^{n}=n\Delta t$ and $t^{n+\frac{1}{2}}=(n+\frac{1}{2})\Delta t$, where $\Delta t$ is the time-step size. The nine independent variables in Eqs.\,(\ref{eq:basic1})-(\ref{eq:basic3}) are discretized on the grid as 
\begin{align*}
\begin{pmatrix}v_{x,i,j,k}^{n+\frac{1}{2}}\\[10pt]
v_{y,i,j,k}^{n+\frac{1}{2}}\\[10pt]
v_{z,i,j,k}^{n+\frac{1}{2}}
\end{pmatrix},\,\begin{pmatrix}E_{x,i+\frac{1}{2},j,k}^{n}\\[10pt]
E_{y,i,j+\frac{1}{2},k}^{n}\\[10pt]
E_{z,i,j,k+\frac{1}{2}}^{n}
\end{pmatrix},\,\begin{pmatrix}B_{x,i,j+\frac{1}{2},k+\frac{1}{2}}^{n+\frac{1}{2}}\\[8pt]
B_{y,i+\frac{1}{2},j,k+\frac{1}{2}}^{n+\frac{1}{2}}\\[8pt]
B_{z,i+\frac{1}{2},j+\frac{1}{2},k}^{n+\frac{1}{2}}
\end{pmatrix},
\end{align*}
where superscript indices are for the time grid, and subscript indices are for the spatial grid. For variables defined on integer grids point $f_{i}$, denote finite difference and average between two adjacent grid points as 
\begin{align*}
\Delta_{x}[f_{i}]:=\dfrac{f_{i+1}-f_{i}}{\Delta x},\quad f_{\,\overline{i+\frac{1}{2}}}:=\dfrac{f_{i+1}+f_{i}}{2}.
\end{align*}
For variables defined on half-integer grid points $g_{i+\frac{1}{2}}$, similarly, 
\begin{align*}
\Delta_{x}\left[g_{i+\frac{1}{2}}\right]:=\dfrac{g_{i+\frac{1}{2}}-g_{i-\frac{1}{2}}}{\Delta x},\quad g_{\,\bar{i}}:=\dfrac{g_{i+\frac{1}{2}}+g_{i-\frac{1}{2}}}{2}.
\end{align*}
Using these notations, the discretization of Eqs.\,(\ref{eq:basic1})-(\ref{eq:basic3}) is 
\begin{equation}
\begin{aligned}\Delta_{t}\left[v_{x,i,j,k}^{n+\frac{1}{2}}\right] & =\omega_{\mathrm{p},i,j,k}\,E_{x,\bar{i},j,k}^{n}-\Omega\,v_{y,i,j,k}^{\bar{n}},\\
\Delta_{t}\left[v_{y,i,j,k}^{n+\frac{1}{2}}\right] & =\omega_{\mathrm{p},i,j,k}\,E_{y,i,\bar{j},k}^{n}+\Omega\,v_{x,i,j,k}^{\bar{n}},\\
\Delta_{t}\left[v_{z,i,j,k}^{n+\frac{1}{2}}\right] & =\omega_{\mathrm{p},i,j,k}\,E_{z,i,j,\bar{k}}^{n}.
\end{aligned}
\label{eq:numerical1}
\end{equation}
\begin{equation}
\begin{aligned}\Delta_{t}\left[E_{x,i+\frac{1}{2},j,k}^{n}\right] & =\Delta_{y}\left[B_{z,i+\frac{1}{2},j+\frac{1}{2},k}^{n+\frac{1}{2}}\right]-\Delta_{z}\left[B_{y,i+\frac{1}{2},j,k+\frac{1}{2}}^{n+\frac{1}{2}}\right]-\left(\omega_{\mathrm{p}}v_{x}^{n+\frac{1}{2}}\right)_{\overline{i+\frac{1}{2}},j,k},\\[10pt]
\Delta_{t}\left[E_{y,i,j+\frac{1}{2},k}^{n}\right] & =\Delta_{z}\left[B_{x,i,j+\frac{1}{2},k+\frac{1}{2}}^{n+\frac{1}{2}}\right]-\Delta_{x}\left[B_{z,i+\frac{1}{2},j+\frac{1}{2},k}^{n+\frac{1}{2}}\right]-\left(\omega_{\mathrm{p}}v_{y}^{n+\frac{1}{2}}\right)_{i,\overline{j+\frac{1}{2}},k},\\[10pt]
\Delta_{t}\left[E_{z,i,j,k+\frac{1}{2}}^{n}\right] & =\Delta_{x}\left[B_{y,i+\frac{1}{2},j,k+\frac{1}{2}}^{n+\frac{1}{2}}\right]-\Delta_{y}\left[B_{x,i,j+\frac{1}{2},k+\frac{1}{2}}^{n+\frac{1}{2}}\right]-\left(\omega_{\mathrm{p}}v_{z}^{n+\frac{1}{2}}\right)_{i,j,\overline{k+\frac{1}{2}}}.
\end{aligned}
\label{eq:numerical2}
\end{equation}
\begin{equation}
\begin{aligned}\Delta_{t}\left[B_{x,i,j+\frac{1}{2},k+\frac{1}{2}}^{n+\frac{1}{2}}\right] & =\Delta_{z}\left[E_{y,i,j+\frac{1}{2},k}^{n}\right]-\Delta_{y}\left[E_{z,i,j,k+\frac{1}{2}}^{n}\right],\\[10pt]
\Delta_{t}\left[B_{y,i+\frac{1}{2},j,k+\frac{1}{2}}^{n+\frac{1}{2}}\right] & =\Delta_{x}\left[E_{z,i,j,k+\frac{1}{2}}^{n}\right]-\Delta_{z}\left[E_{x,i+\frac{1}{2},j,k}^{n}\right],\\[10pt]
\Delta_{t}\left[B_{z,i+\frac{1}{2},j+\frac{1}{2},k}^{n+\frac{1}{2}}\right] & =\Delta_{y}\left[E_{x,i+\frac{1}{2},j,k}^{n}\right]-\Delta_{x}\left[E_{y,i,j+\frac{1}{2},k}^{n}\right].
\end{aligned}
\label{eq:numerical3}
\end{equation}
Here, $\omega_{\mathrm{p}}(\boldsymbol{r})$ is a prescribed function, $\omega_{\mathrm{p},i,j,k}:=\omega_{\mathrm{p}}(x_{i},y_{j},z_{k})$, ${\displaystyle \left(\omega_{\mathrm{p}}v_{x}\right)_{\overline{i+\frac{1}{2}}}}:=(\omega_{\mathrm{p},i}v_{x,i}+\omega_{\mathrm{p},i+1}v_{x,i+1})/2$. Notice that although its right-hand side depends on $v^{n+\frac{1}{2}}$ through $v^{\bar{n}}$, Eq.\,(\ref{eq:numerical1}) is explicitly solvable by the Cayley transformation that appears in the Boris algorithm \citep{qin2013boris} and other structure-preserving algorithm for charged particle dynamics \citep{he2015volume,He16Higher,He2016HigherRela,zhang2015volume,Fu2022}.

The algorithm in 2D is a special case of the 3D algorithm described above. When the plasma density is invariant along the background magnetic field, $\omega_{\mathrm{p}}(\boldsymbol{r})=\omega_{\mathrm{p}}(x,y)$, we can replace the $z$ dependence of all variables by $\exp(\mathrm{i}k_{z}z)$. Then, all variables in Eqs.\,(\ref{eq:numerical1})-(\ref{eq:numerical3}) no longer depend on the index $k$, and a constant $\mathrm{i}k_{z}$ replaces the difference operators $\Delta_{z}[\cdot]$.

\subsection{Propagation of the TLCW in 2D and 3D}

We now apply the algorithm to simulate the TLCW excited by a localized time-dependent source in two- and three-dimensions. An external force that represents the source is added to the RHS of the momentum equation (\ref{eq:numerical1}). In the 3D simulation, the force we used is 
\begin{align}
\boldsymbol{F}(\boldsymbol{r},t)\sim\exp\left(-\dfrac{|\boldsymbol{r}-\boldsymbol{r}_{s}|^{2}}{\delta^{2}}\right)e^{\mathrm{i}(k_{s}z-\omega_{s}t)},
\end{align}
where $\boldsymbol{r}_{s}$ and $\delta$ are the spatial center and width of the source, $k_{s}$ and $\omega_{s}$ are the wavenumber and frequency of the source. The direction of the force does not affect the excitation of the surface waves. In the 2D simulation, the force does not depend on $z$, i.e., $k_{s}=0,$ and the factor $\mathrm{i}k_{z}$ replaces the difference operators $\Delta_{z}[\cdot]$ in Eqs.\,(\ref{eq:numerical2}) and (\ref{eq:numerical3}).

\begin{figure}
\centering %
\begin{minipage}[c]{6cm}%
\centering 
\includegraphics[width=6cm]{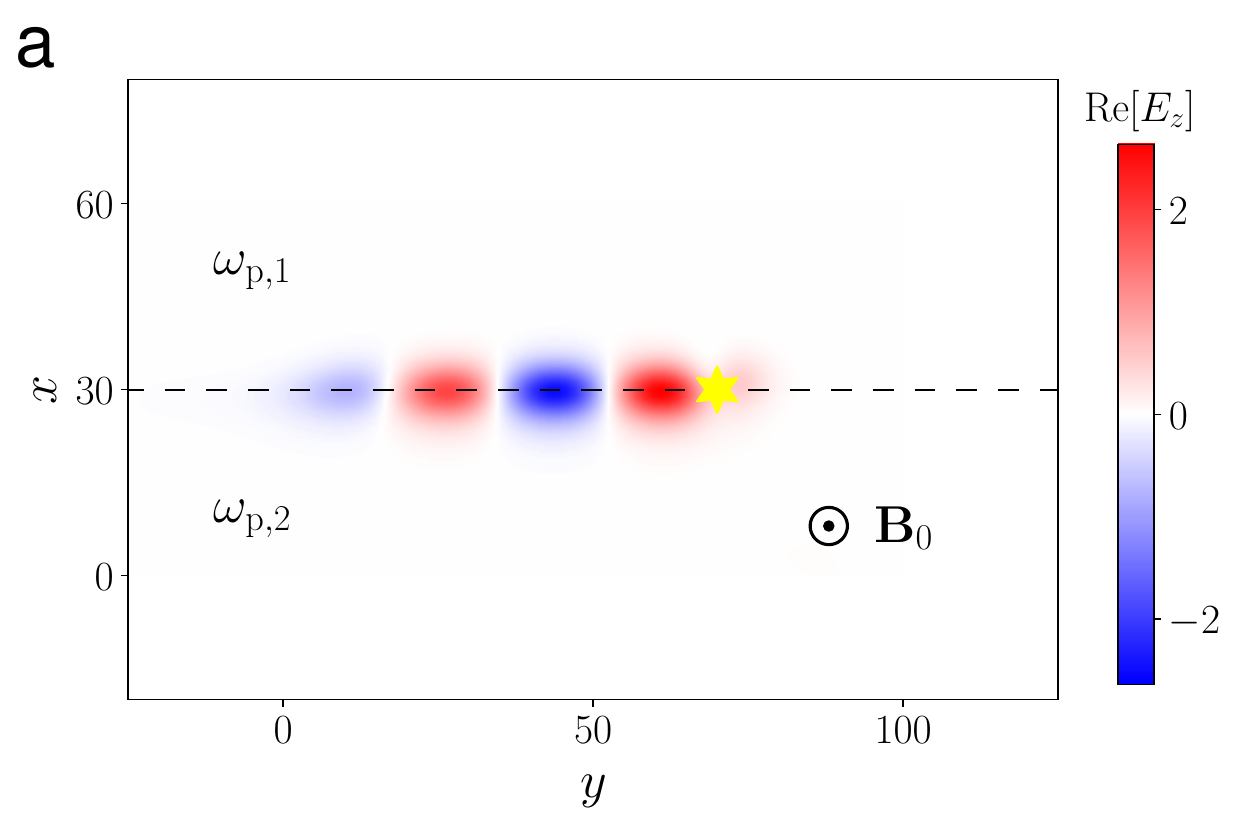}%
\end{minipage}
\hspace{0.5cm} %
\begin{minipage}[c]{3.6cm}%
\centering 
\includegraphics[width=3.6cm]{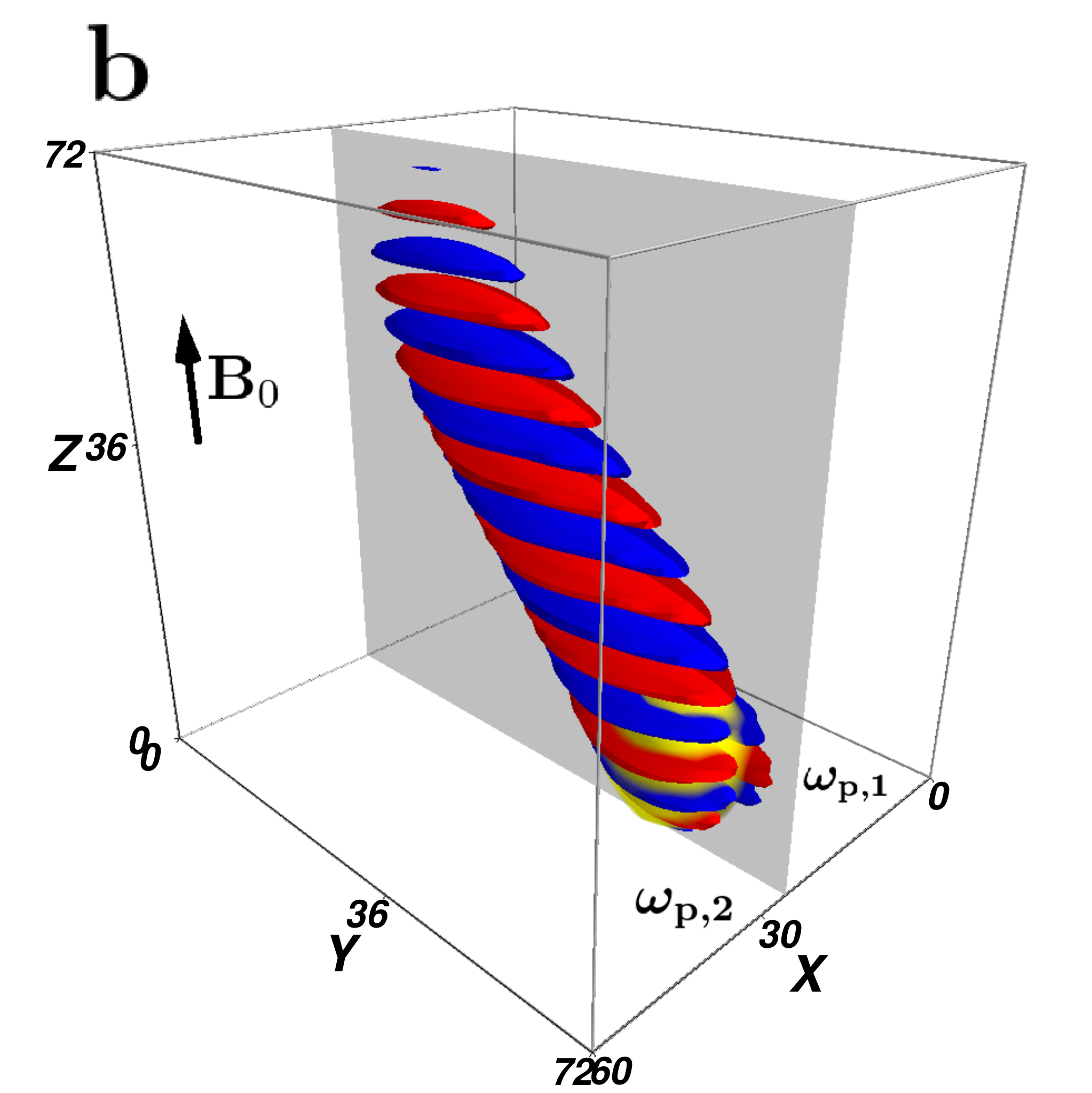}%
\end{minipage}
\caption{Propagation of TLCW in (a) 2D and (b) 3D at $|t\Omega|=250$, where $\omega_{\mathrm{p,1}}=0.8$, $\omega_{\mathrm{p,2}}=0.3$, $k_{s}=0.8$, and $\omega_{s}=0.5$. In (a), the color map indicates the strength of real part of $E_{z}$, the yellow star marks the location of the source, and the dashed line is the interface between two regions. In (b), the red and blue contours represent the locations of $\mathrm{Re}[E_{z}]=\pm0.15\,\mathrm{max}[\mathrm{Re}(E_{z})]$, the yellow sphere marks the source, and the gray surface is the interface between two regions.}
\label{fig:stripe} 
\end{figure}

The parameters for the first case simulated is $\omega_{\mathrm{p,1}}=0.8$, $\omega_{\mathrm{p,2}}=0.3$, $k_{s}=0.8$, and $\omega_{s}=0.5$. The dispersion relation calculated in 1D was shown in Figs.\,\ref{fig:isofreq_nontrivial}a and \ref{fig:isofreq_nontrivial}c. The center of the source is on the interface. The time-dependent propagation of the TLCW in 2D and 3D after a finite time are shown in Fig.\,\ref{fig:stripe}. Notice that the configuration is equivalent to the right interface shown in Fig.\,\ref{fig:slab_geometry}, where the blue lines in Fig.\,\ref{fig:isofreq_nontrivial} represent the surface waves in Fig.\,\ref{fig:isofreq_nontrivial}. In the 2D simulation displayed in Fig.\,\ref{fig:stripe}a, the TLCW is excited, and the excitation of bulk waves is negligible. Furthermore, the surface wave only travels to the left side of the source, confirming its unidirectional propagation. The same phenomenons can also be observed in 3D simulation shown in Fig.\,\ref{fig:stripe}b, where the TLCW is excited and propagates towards the upper-left direction in the $y$-$z$ plane.

\subsection{The momentum of the surface waves}

In this subsection, we discuss the momentum and angular momentum carried by the TLCW. There exist two different definitions of momentum for plasma waves \citep{barnett2010resolution,dodin2012axiomatic}, i.e., the (Minkowski) canonical momentum and the (Abraham) kinetic momentum. The canonical momentum of the TLCW is proportional to the wavenumber $\boldsymbol{k}$, and it can vary significantly, as illustrated in Fig.\,\ref{fig:isofreq_nontrivial}. However, the kinetic momentum, which is proportional to it group velocity $\boldsymbol{v}_{\mathrm{g}}$, is unidirectional. For example, the surface waves on the right interface can only have kinetic momentum in the upper-left or lower-left direction in the $y$-$z$ plane.

\begin{figure}
\centering %
\begin{minipage}[c]{3.75cm}%
\includegraphics[width=3.75cm]{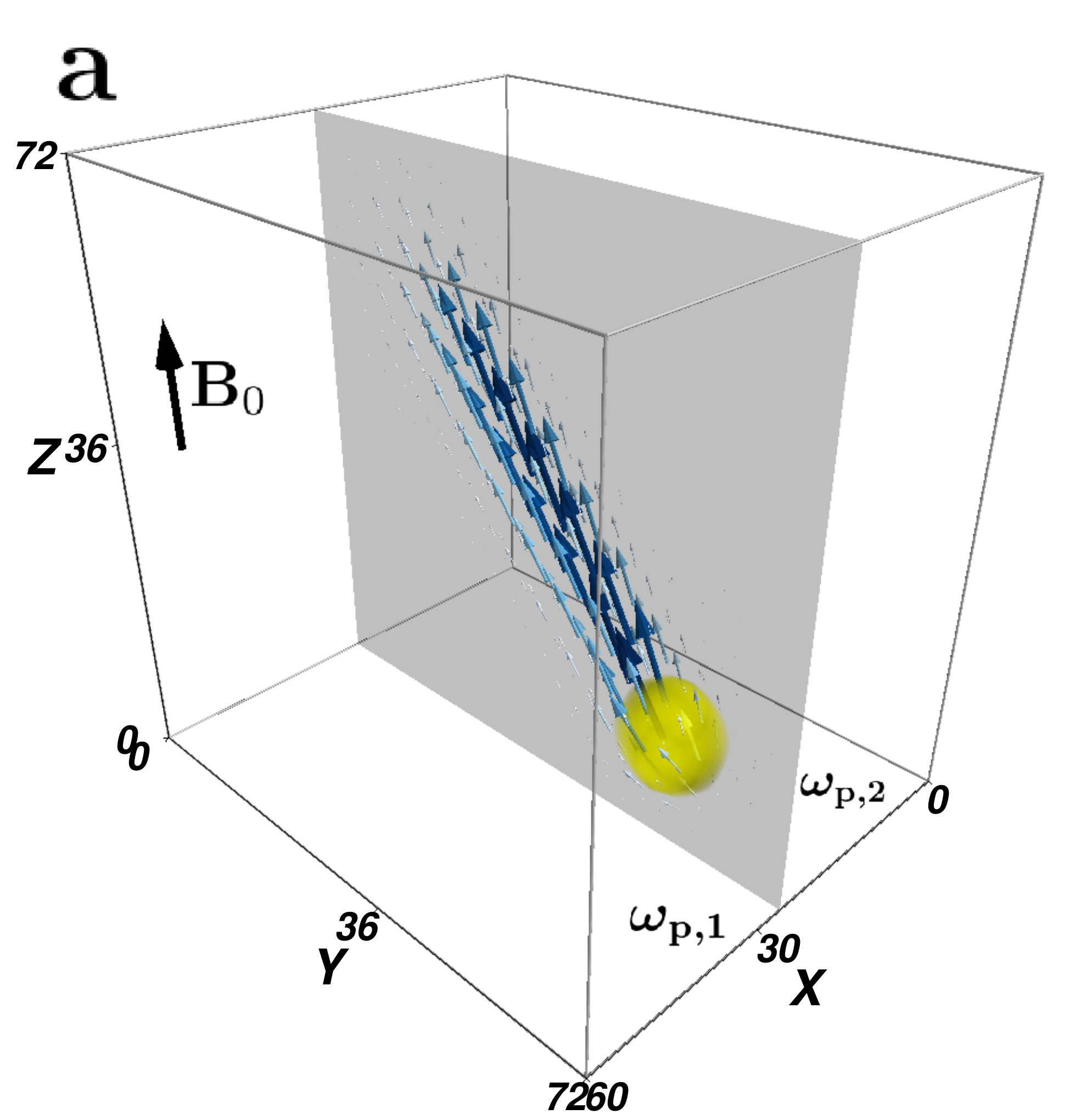}%
\end{minipage}\hspace{0.25cm} %
\begin{minipage}[c]{4.25cm}%
\includegraphics[width=4.25cm]{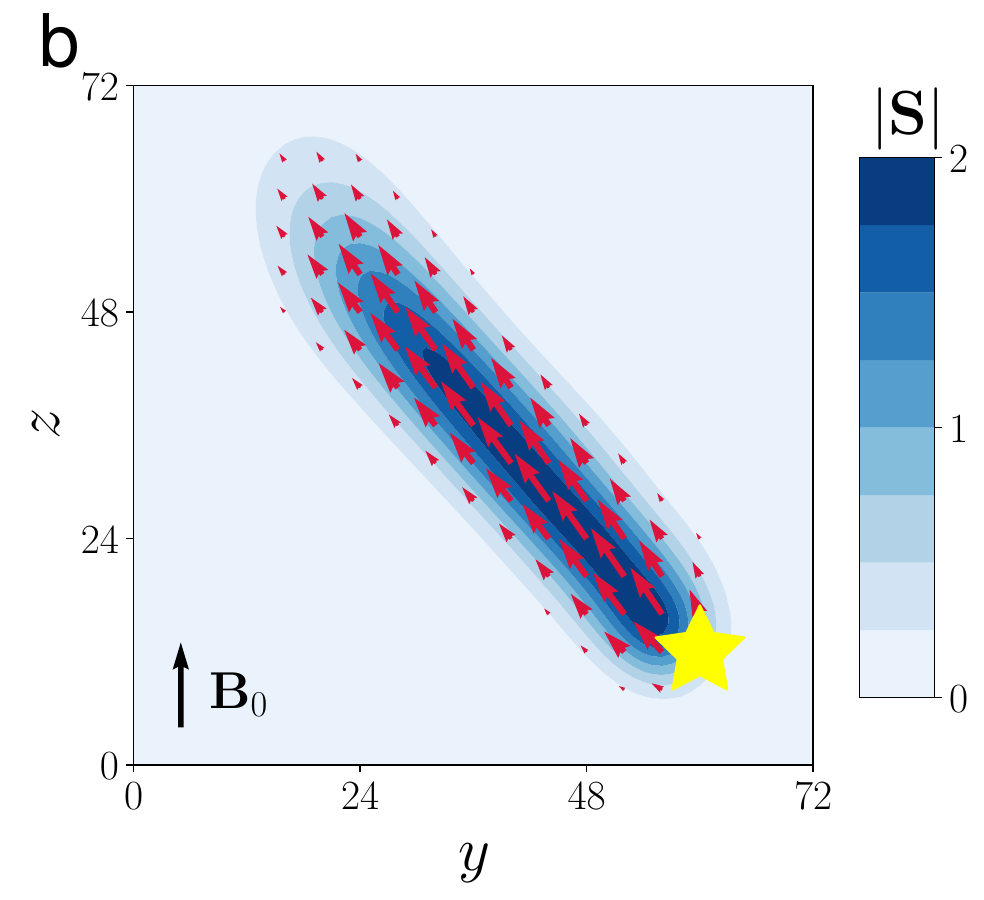}%
\end{minipage}\caption{(a) The Poynting vectors in the 3D simulation. The yellow sphere marks the source, the arrows indicate the Poynting vectors, the gray surface is the interface. (b) The Poynting vectors at the interface between two regions (at $x=30$). The yellow star marks the source, the red arrows indicate the Poynting vectors in the $y$-$z$ plane, whose magnitudes are given by the color of the background.}
\label{fig:poynting} 
\end{figure}

For the cold plasma model given by Eqs.\,(\ref{eq:basic1})-(\ref{eq:basic3}), the dielectric tensor is Hermitian and independent of $\boldsymbol{k}$. Therefore, the kinetic momentum, the group velocity, and the flux of canonical energy are all proportional to the Poynting vectors \citep{stix1992waves} $\boldsymbol{S}\sim\mathrm{Re}[\boldsymbol{E}\times\boldsymbol{B}^{*}]$. The Poynting vectors of the TLCW excited on a planar interface is shown in Fig.\,\ref{fig:poynting}. As expected, the kinetic momentum is along the direction of wave propagation

\subsection{Propagation of the TLCW on non-planar interface}

\begin{figure}
\centering %
\begin{minipage}[c]{5.25cm}%
\includegraphics[width=5.25cm]{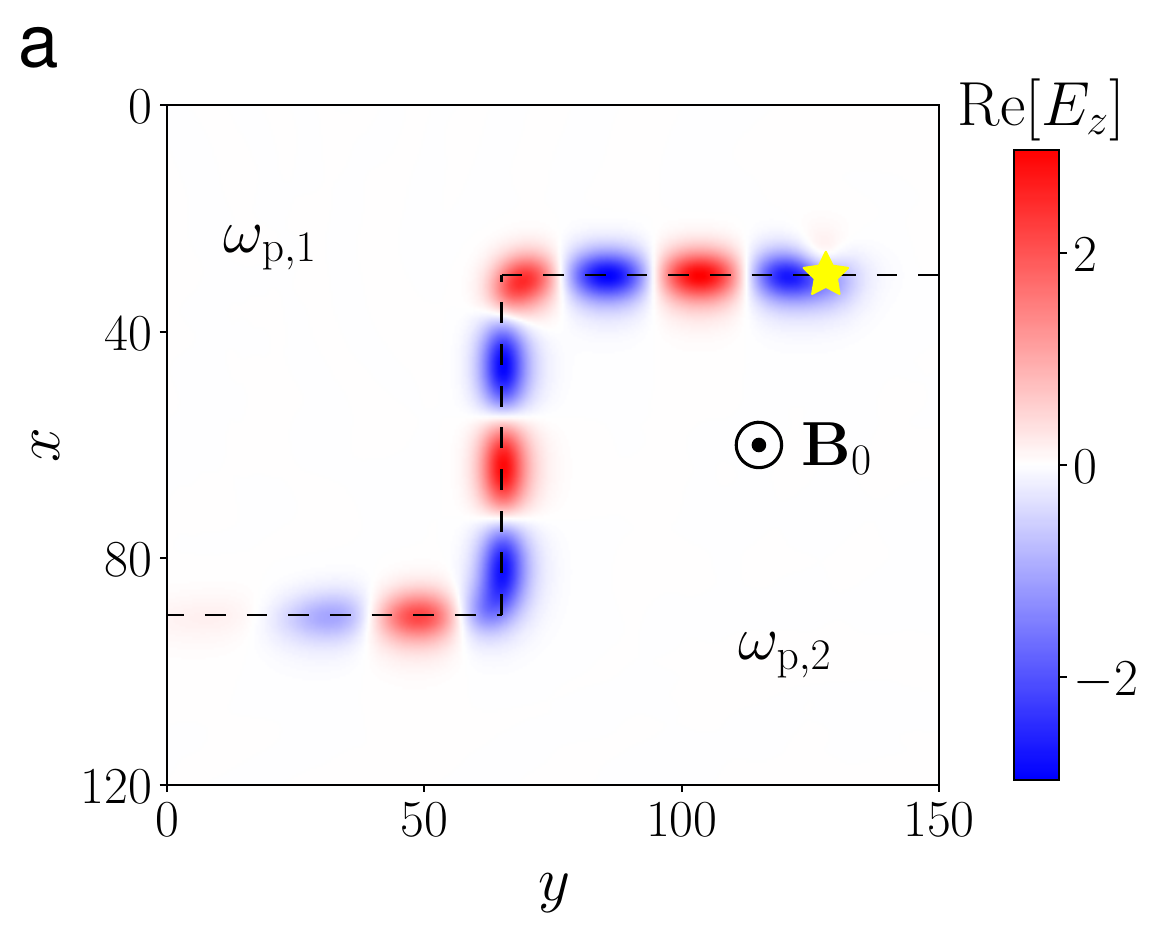}%
\end{minipage}\hspace{0.25cm} %
\begin{minipage}[c]{3.5cm}%
\includegraphics[width=3.5cm]{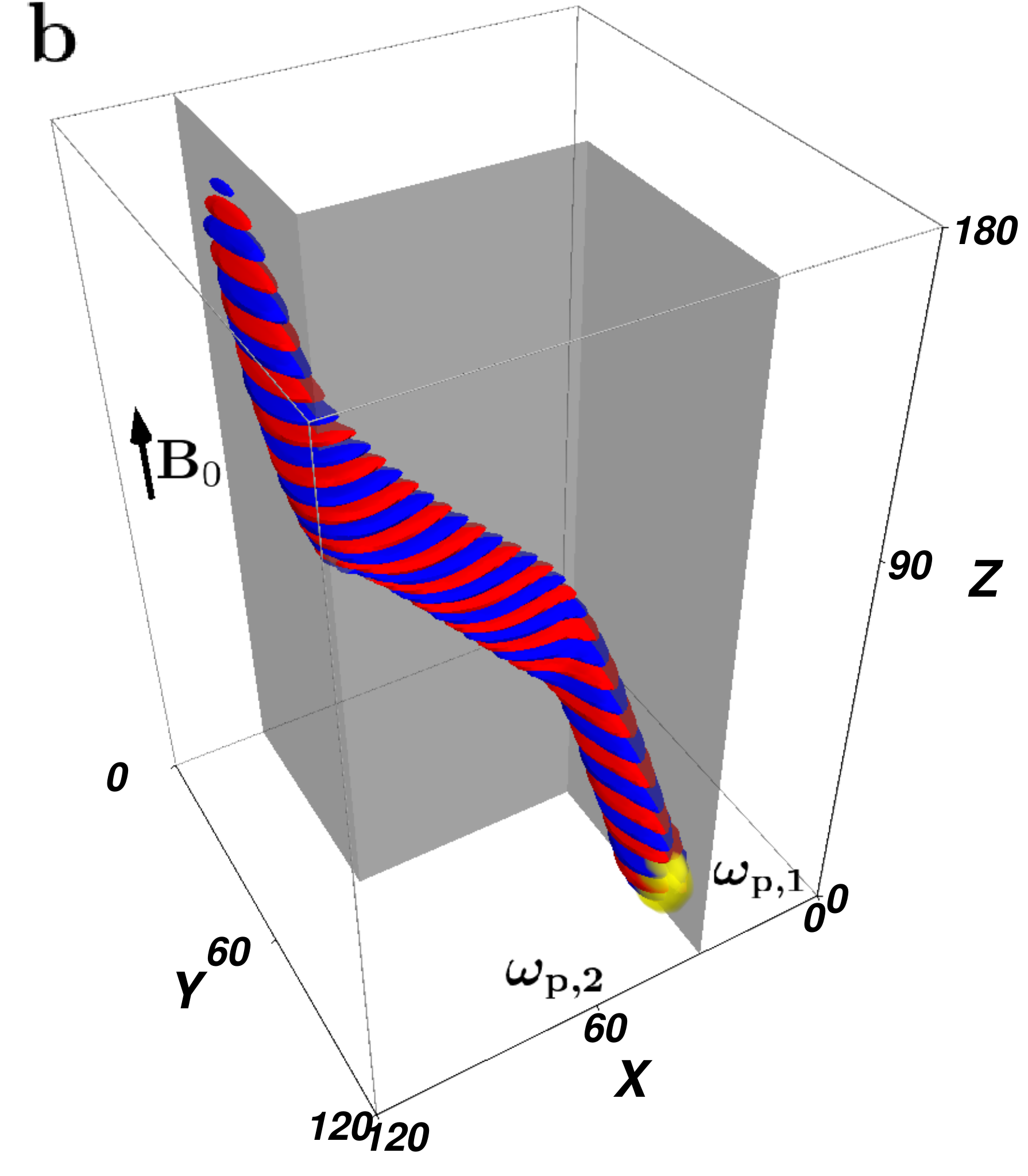}%
\end{minipage}\caption{(a) 2D and (b) 3D simulations of the TLCW excited at a zigzagged interface The source was turned on at $t=0$, and the field strength was plotted at $|t\Omega|=600$.}
\label{fig:zigzag} 
\end{figure}

The main properties of the TLCW is robust in more complex configurations. We now show some simulation results of the TLCW on non-planar boundaries in both two- and three-dimensions. Shown in Fig.\,\ref{fig:zigzag} is the TLCW propagation on a zigzagged interface after a finite time, which clearly demonstrates immunity of back-scattering and unidirectional propagation. When the TLCW meets sharp turns on the interface, it propagates along the interface without reflection or transition to bulk waves. 

\begin{figure}
\centering %
\begin{minipage}[c]{5.25cm}%
\includegraphics[width=5.25cm]{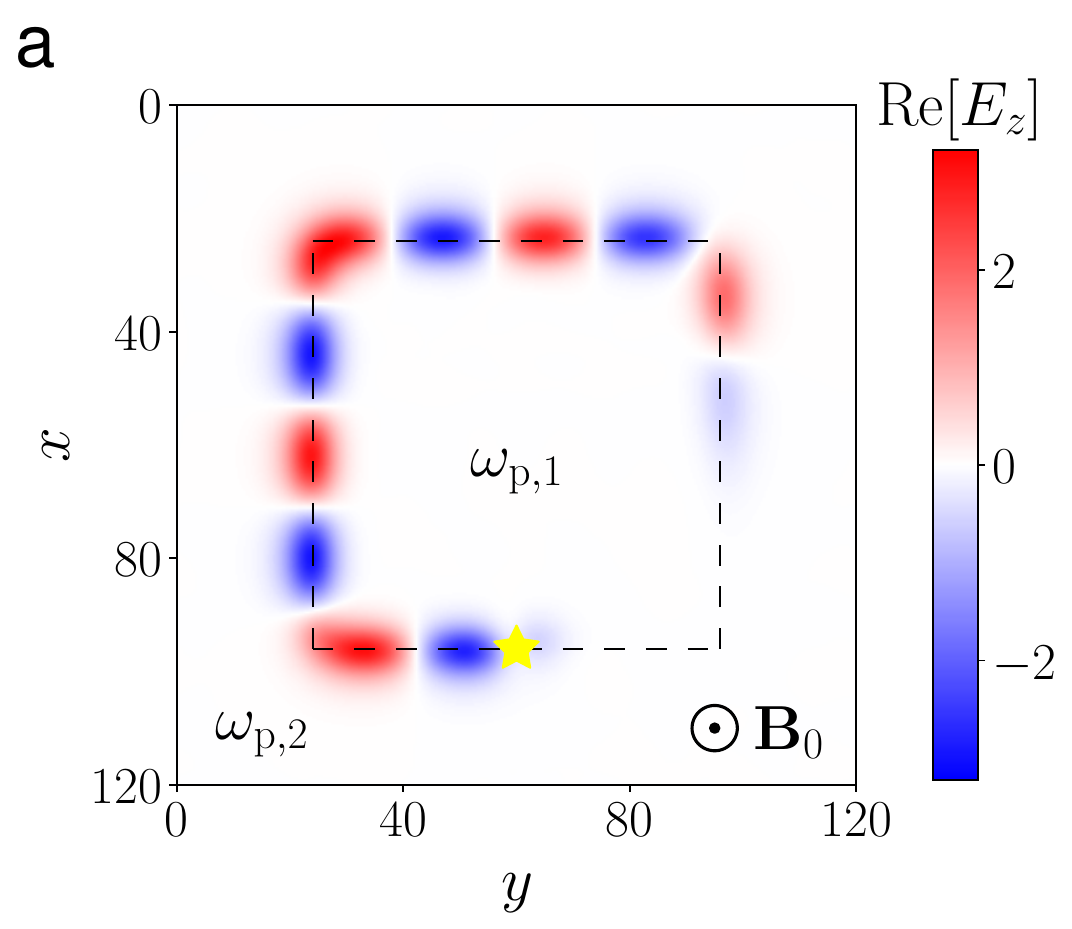}%
\end{minipage}\hspace{0.25cm} %
\begin{minipage}[c]{3cm}%
\includegraphics[width=3cm]{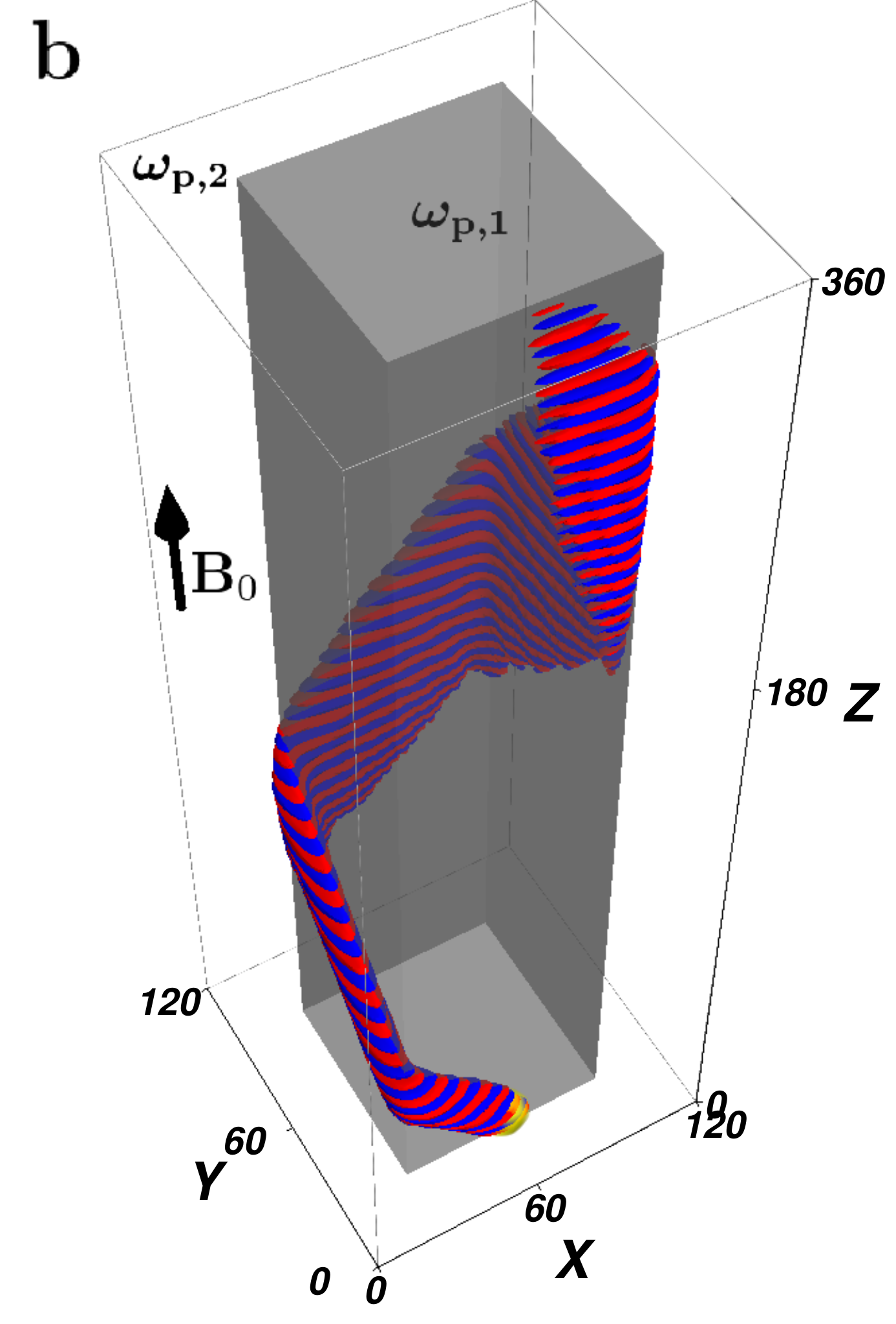}%
\end{minipage}\caption{(a) 2D and (b) 3D simulations of the TLCW excited on a square interface. The source was turned on at $t=0$, and the field strength was plotted at $|t\Omega|=600$.}
\label{fig:square} 
\end{figure}

\begin{figure}
\centering %
\begin{minipage}[c]{5.25cm}%
\includegraphics[width=5.25cm]{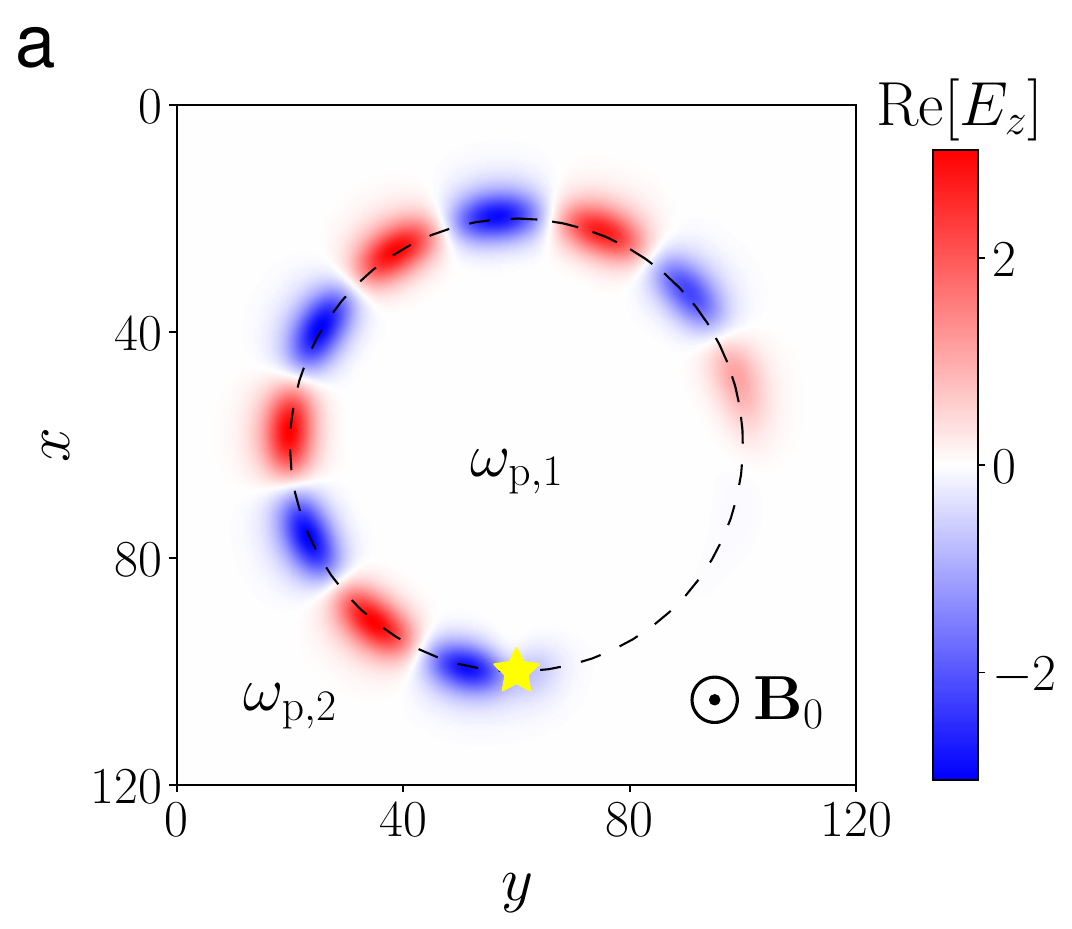}%
\end{minipage}\hspace{0.25cm} %
\begin{minipage}[c]{3cm}%
\includegraphics[width=3cm]{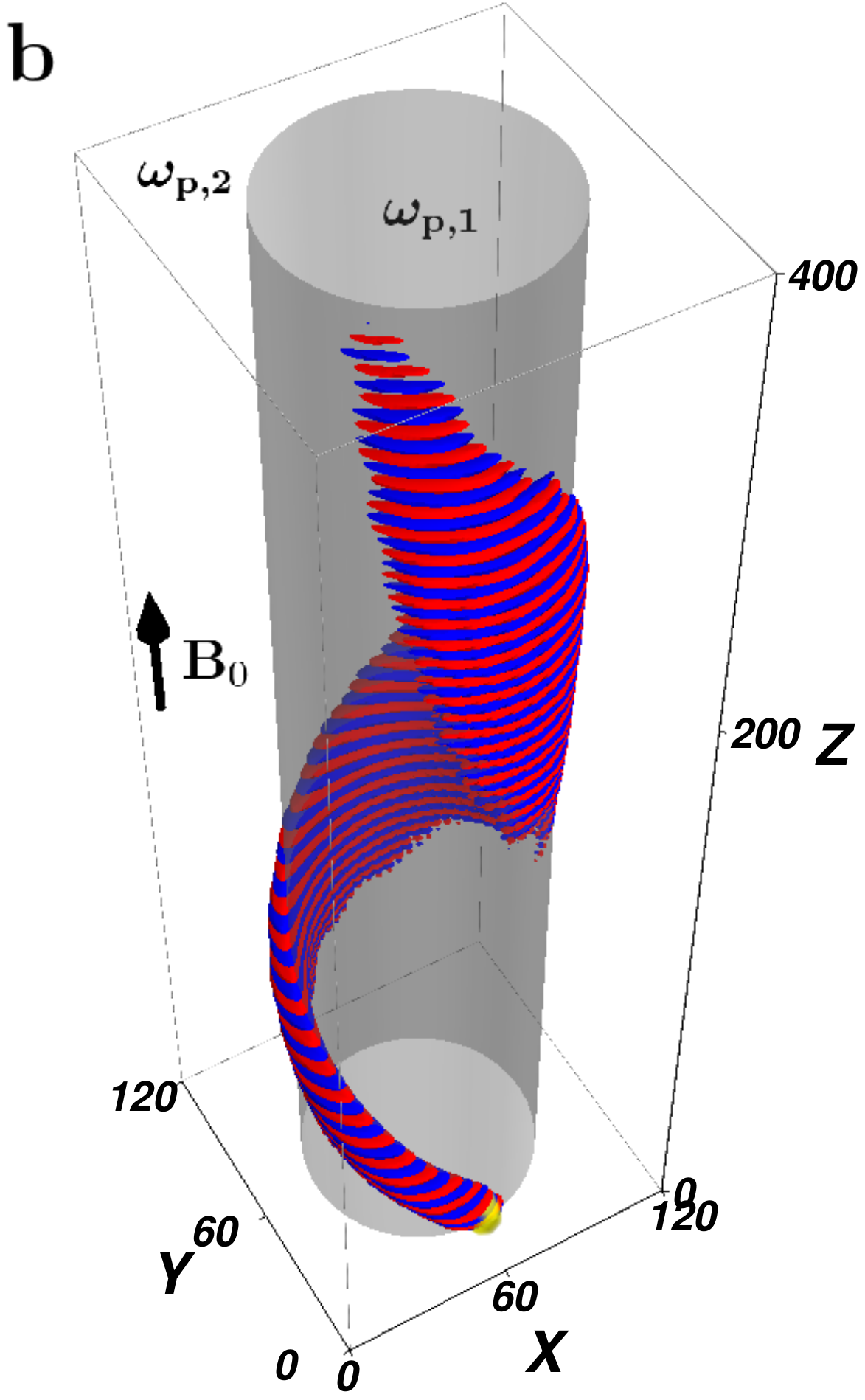}%
\end{minipage}\caption{(a) 2D and (b) 3D simulations of the TLCW excited on a circular interface. The source was turned on at $t=0$, and the field strength was plotted at $|t\Omega|=600$.}
\label{fig:cylinder} 
\end{figure}

Additional examples are shown in Figs.\,\ref{fig:square} and \ref{fig:cylinder}, where the interface between two regions is a square or a circle. In these configurations, the left and right boundaries are connected and indistinguishable. As anticipated, the TLCW propagates along the boundaries, regardless their shapes, in a unidirectional manner and without any scattering. When viewed against the direction of the magnetic field, the TLCW propagates clockwise if $\omega_{\mathrm{p,1}}>\omega_{\mathrm{p,2}}$, and therefore carries a non-zero kinetic angular momentum proportional to $\int\boldsymbol{r}\times\boldsymbol{S}\,\mathrm{d}V$. Notice that the source exciting the TLCW does not carry any angular momentum. The existence of an angular-momentum-carrying surface wave reflects the topological property of the bulk material and is predicted by the bulk-edge correspondence. Because of these desirable properties, the TLCW could be explored as an effective mechanism for driving current and flow in magnetized plasmas.

\section{Conclusions and discussion\label{sec:conclusion}}

We have studied the dispersion relation and propagation of the recently identified Topological Langmuir-Cyclotron Wave (TLCW). From the excitation condition (\ref{eq:condition}), we derived the frequency range in Eq.\,(\ref{eq:fre_range}) where the TLCW can be observed. It is shown that the topological phase transition responsible for the TLCW excitation occurs at the resonance between the Langmuir wave and cyclotron wave, and concurrently with the transition of the shapes of the index-of-refraction surfaces in the familiar CMA diagram.

Based on the isofrequency contours, the group velocity of the TLCW is unidirectional, and so is its kinetic momentum. Furthermore, the TLCW is immune to back-scattering because they lie in the frequency gap of bulk waves. The unidirectional propagation and immunity of back-scattering are verified using time-dependent simulations in two- and three-dimensions. The excitation of the TLCW could be an effective mechanism for driving current and flow in magnetized plasmas.

In the present study, we theoretically investigated the dispersion and propagation of the TLCW using a linearized cold plasma model. For experimental observation, realistic effects need to be carefully considered to better understand the properties and applications of the TLCW. For example, the kinetic effects of wave-particle interaction will play an essential role in the deposition of momentum carried by the TLCW. Other non-Hermitian \citep{bergholtz2021exceptional} and non-linear \citep{smirnova2020nonlinear,zhou2020topological,bergholtz2021exceptional} effects may also be important for topological waves in more complex fluid and plasma models \citep{qin2019kelvin,fu2020physics,qin2021spontaneous,zhu2021topology}.

\section*{Supplementary data}
Supplementary movies are available at (DOI).

\section*{Declaration of interests}
The authors report no conflict of interest.

\section*{Funding}
This research was supported by the U.S. Department of Energy (DE-AC02-09CH11466). 

\section*{Author ORCID}
Y. Fu, https://orcid.org/0000-0002-5205-9827. H. Qin, https://orcid.org/0000-0003-0304-3762.

\bibliographystyle{jpp}
\bibliography{ref}

\end{document}